\documentclass[3p,times,twocolumn]{elsarticle}
 \biboptions{comma,sort&compress}
 
\usepackage{graphicx}
\usepackage{here}
\usepackage{ecrc}

\volume{00}

\firstpage{1}

\journalname{Nuclear and Particle Physics Proceedings}

\runauth{F. Giuli}

\jid{nppp}

\jnltitlelogo{Nuclear and Particle Physics Proceedings}

\usepackage{amssymb}
 \usepackage{lineno}

\usepackage[figuresright]{rotating}

\begin{document}

\begin{frontmatter}

%%
%%%%%%%%%%%%%%%%%%%%%%%%%%%%%%%%%%%%%%%%%%%%%%%%%
\title{ Impact of ATLAS $V$ + jets data on PDF fits
 $^*$}
 % \corref{cor0}}
 \cortext[cor0]{Talk given at 23rd International Conference in Quantum Chromodynamics (QCD 20,  35th anniversary),  27 - 30 October 2020, Montpellier - FR}
 \author[label1,label2]{F. Giuli}

\ead{francesco.giuli@roma2.infn.it, francesco.giuli@cern.ch}
\address[label1]{University and INFN Rome Tor Vergata, Via della Ricerca Scientifica 1, 00133, Rome, Italy}
  \fntext[label2]{On behalf of the ATLAS Collaboration - Proceeding licensed with copyright CERN for the benefit of the ATLAS Collaboration. CC-BY-4.0 license}
  %\fntext[label3]{Proceeding licensed with copyright CERN for the benefit of the ATLAS Collaboration. CC-BY-4.0 license}

\pagestyle{myheadings}
\markright{ }
\begin{abstract}
This proceeding presents a new set of proton parton distribution functions, ATLASepWZVjet20, produced in an analysis at next-to-next-to-leading-order in QCD. The new datasets considered are the ATLAS measurements of $W^{\pm}$ and $Z$ boson production in association with jets in $pp$ collisions at $\sqrt{s}$ = 8 TeV at the LHC with integrated luminosities of 20.2 fb$^{-1}$ and 19.9 fb$^{-1}$  respectively. The analysis also considers the ATLAS measurements of differential $W^{\pm}$ and $Z$ boson production at $\sqrt{s}$ = 7 TeV with an integrated luminosity of 4.6 fb$^{-1}$ and deep-inelastic scattering data from $e^{\pm}p$ collisions at the HERA accelerator. An improved determination of the sea-quark densities at high Bjorken $x$ is shown, while confirming a strange-quark density of similar size as the up-and down-sea quark densities in the range $x\sim$ 0.02 found by previous ATLAS analyses.
\end{abstract}
% \begin{document}
\begin{keyword}  
%% keywords here, in the form: keyword \sep keyword
ATLAS \sep SM \sep PDF \sep $V$ + jets
%% MSC codes here, in the form: \MSC code \sep code
%% or \MSC[2008] code \sep code (2000 is the default)

\end{keyword}

\end{frontmatter}

%\linenumbers

%%%%%%%%%%%%
%\vspace*{-1.5cm}
\section{Input datasets}

The final combined $e^{\pm}p$ cross section measurements at HERA~\cite{herapdf20} cover the kinematic range of Bjorken $x$ from 0.65 down to 6 $\cdot$ 10$^{-7}$ and of $Q^{2}$ from 0.045 GeV$^{2}$ to 50 TeV$^{2}$. For this data set, there are 169 correlated sources of uncertainty, ranging from 1.5\% to 3\%.\\
On top of the above described data, several datasets recorded by the ATLAS detector~\cite{Aad:2008zzm} at the LHC~\cite{LHC} are considered.
The ATLAS $W,Z$ differential cross sections are based on data recorded during $pp$ collisions with $\sqrt{s}$ = 7 TeV, and a total integrated luminosity of 4.6 fb$^{-1}$, in the electron and muon boson decay channels~\cite{STDM-2012-20}. There are a total of 131 sources of correlated systematic uncertainty across the $W$ and $Z$ data sets, and these data were used before the combination of the measurements of the electron and muon decay channels. This choice was made to correlate sources of correlated uncertainties to other data sets.\\
The ATLAS $W^{\pm}$ + jets differential cross sections are based on data recorded during $pp$ collisions with $\sqrt{s}$ = 8 TeV and a total integrated luminosity of 20.2 fb$^{-1}$,  in the electron decay channel only~\cite{STDM-2016-14}. The experimental uncertainty ranges from 8.2\% to 22.1\% and there are 53 sources of correlated systematic uncertainty in common between the $W^{+}$ and $W^{-}$ spectra.\\
The ATLAS $Z$ + jets double-differential cross sections are also based on data recorded during $pp$ collisions with $\sqrt{s}$ = 8 TeV and a total integrated luminosity of 19.9 fb$^{-1}$, in the $Z\rightarrow ee$ decay channel~\cite{STDM-2016-11}. The experimental precision ranges from 4.7\% to 37.1\%. There are 42 sources of correlated systematic uncertainty and two sources of uncorrelated uncertainty relating to the data and background Monte Carlo (MC) simulation statistics.

\section{Fit framework}
The determination of proton PDFs uses the \texttt{xFitter} framework~\cite{herapdf20,HERAFitter,Aaron:2009kv}. The DGLAP evolution equations of QCD yield the proton PDFs at any value of $Q^2$. PDFs are parametrised as functions of $x$ at an initial scale $Q^2_0$ = 1.9~GeV$^2$, such that it is below the charm-mass matching scale, $\mu_{c}^{2}$, which is set equal to the charm mass, $\mu_{c} = m_{c}$. 
The heavy quark masses are set to their pole masses as determined by a combined analysis of HERA data on inclusive and heavy-flavour DIS processes~\cite{herapdf20,H1:2018flt}, $m_c$ = 1.43 GeV and $m_b$ = 4.5 GeV, and the strong coupling constant is fixed to $\alpha_S(m_Z)$ =  0.118. These choices follow those of the HERAPDF2.0 fit~\cite{herapdf20}.\\
The quark distributions at the initial scale are assumed to behave according to the following parametrisation also used by the HERAPDF2.0 and ATLASepWZ16 fits~\cite{herapdf20,STDM-2012-20} 
\begin{equation}
 xq_i(x) = A_i x^{B_i} (1-x)^{C_i} P_i(x)\, ,
\label{eqn:pdf}
\end{equation}
where $P_i(x)= (1 + D_i x +E_i x^2) e^{F_ix}$. The parametrised quark distributions, $xq_i$, are chosen to be the valence quark distributions ($xu_v,~xd_v$) and the light anti-quark distributions ($x\bar{u},~x\bar{d},~x\bar{s}$). The gluon distribution is parametrised with the more flexible form
\begin{equation}
xg(x) = A_g x^{B_g} (1-x)^{C_g}P_g(x)  - A'_g x^{B'_g} (1-x)^{C'_g},
\end{equation}
where $C'_g$ is fixed to a value of $25$ to suppress negative contributions of the primed term at high $x$, as in Ref.~\cite{0901.0002}.\\
The $D,E$ and $F$ terms in the expression $P_i(x)$ are used only if required by the data, following the procedure described in Ref.~\cite{herapdf20}.  For the ATLASepWZVjet20 fit~\cite{CONF_2020}, this results in the usage of two additional parameters: $E_{u_v}$ and $D_{g}$.  There is a total of 16 free parameters used in the central fit. The agreement of the data with the fit predictions is quantified with a $\chi^2$. The definition of the $\chi^2$  is the same as adopted in Refs.~\cite{herapdf20,STDM-2012-20}.

\section{Results}

Throughout this section, the presented ATLASepWZVjet20 PDFs are compared to an equivalent fit without the $V$ + jets data, labelled ATLASepWZ20.

\subsection{Goodness of the fit}
\label{subsec:DataFits}
\begin{figure}[t!]
  \centering
  \includegraphics[width=.9\linewidth]{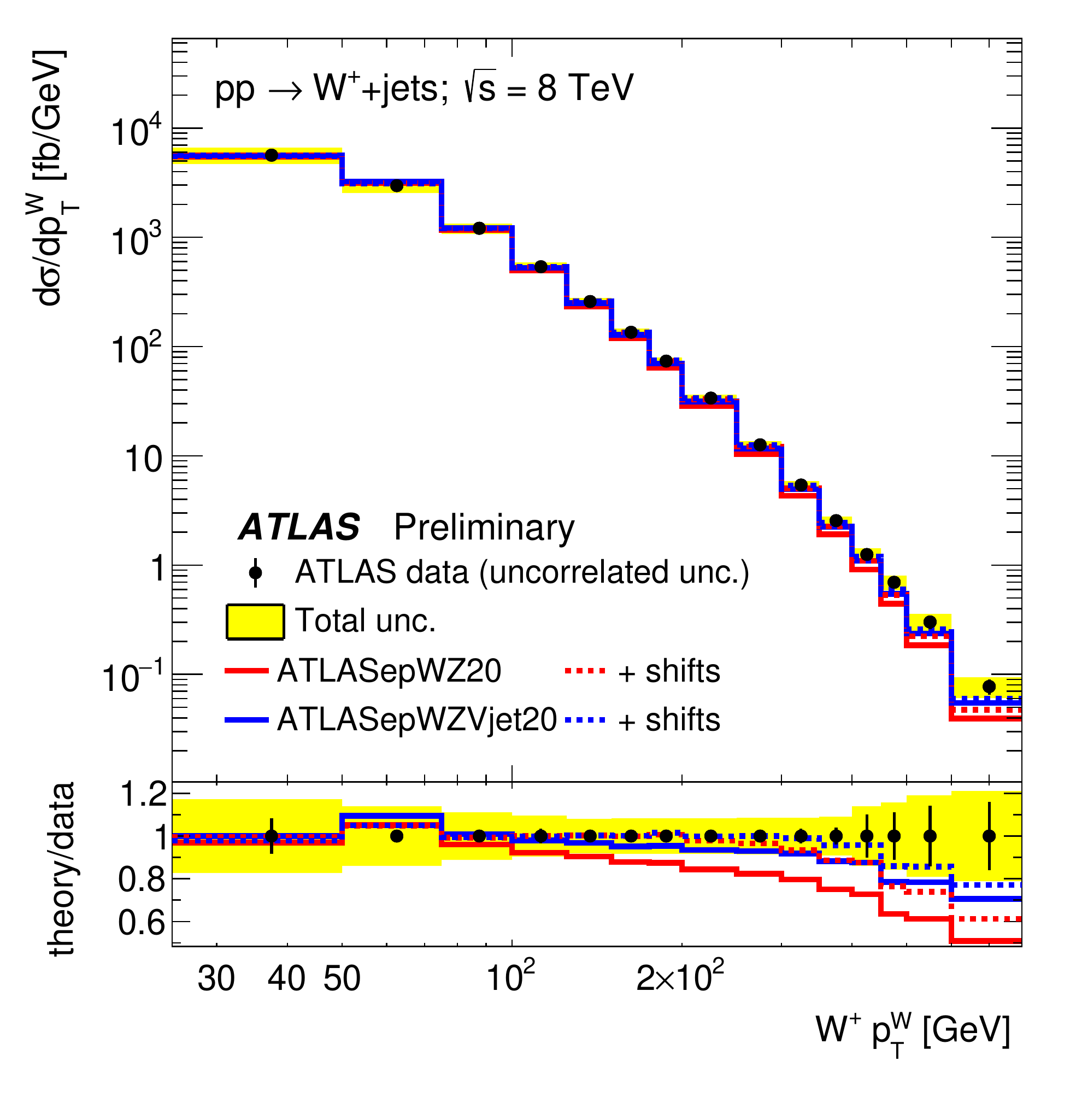}
	\caption{The differential cross sections of $W^+$ + jets as a function of the transverse momentum of the $W$ boson, $p_{\mathrm{T}}^{W}$. The bin-to-bin uncorrelated part of the data uncertainties is shown as black error bars, while the total uncertainties are shown as a yellow band. The cross sections are compared to the predictions computed with the predetermined PDFs resulting from the fits ATLASepWZ20 (red lines) and ATLASepWZVjet20 (blue lines). The solid lines show the predictions without shifts of the systematic uncertainties, while for the dashed lines the $b_j$ parameters associated with the experimental systematic uncertainties are allowed to vary to minimise the $\chi^2$. This plot is taken from Ref.~\cite{CONF_2020}.}
  \label{fig:DataPlots1}
\end{figure}

The total $\chi^2$ per degree of freedom (NDF) for the ATLASepWZVjet20 fit, along with the partial $\chi^2$ per data point (NDP) and correlated $\chi^2$ for each data set entering the fit, is given in Table~\ref{tab:chi}. The partial $\chi^2$ for the HERA and ATLAS inclusive $W, Z$ data in fits including the additional data are similar to those obtained in fits without these additional data, demonstrating that there is no tension between these sets. The partial $\chi^2$ of the $W$ + jets and $Z$ + jets data is reasonable, and neither the HERA nor ATLAS correlated $\chi^2$ is observed to increase significantly with the inclusion of this data.
\begin{table}[h!]\small
 \centering
\begin{tabular}{c c}
\hline
	Fit & ATLASepWZVjet20\\
	\hline
Total $\chi^2/\mathrm{NDF}$ & 1460 / 1198\\
HERA partial $\chi^2/\mathrm{NDP}$ & 1132 / 1016\\
HERA correlated $\chi^2$ & 50\\
HERA log penalty $\chi^2$ & $-$12\\
ATLAS $W, Z$ partial $\chi^2/\mathrm{NDP}$ & 113/105\\
ATLAS $W$ + jets partial $\chi^2/\mathrm{NDP}$ & 25/30\\
ATLAS $Z$ + jets partial $\chi^2/\mathrm{NDP}$ & 82/63\\
ATLAS correlated $\chi^2$ & 65\\
ATLAS log penalty $\chi^2$ & 6\\
\hline
\end{tabular}
\caption{The $\chi^2$ values split by partial, correlated and log penalty for each data set entering the ATLASepWZVjet20 fit. The partial component of the $\chi^2$ for each data set is shown compared to the number of data points of that data set (NDP). This table is taken from Ref.~\cite{CONF_2020}.}
\label{tab:chi}
\end{table}
Figure~\ref{fig:DataPlots1} shows a comparison of the $W^+$ + jets differential cross section measurements with the predictions for each of the ATLASepWZ20 and ATLASepWZVjet20 fits. Adding the $V$ + jets data to the fit improves the $W$ + jets description significantly, particularly in the $W^+$ spectrum where the agreement improves by approximately 20\% at high $p_{\mathrm{T}}^{W}$. A similar (but less significant) improvement has also been found when considering the $Z$ + jets differential cross section measurements.

\subsection{Resulting PDFs}
\begin{figure}
  \centering
\includegraphics[width=.9\linewidth]{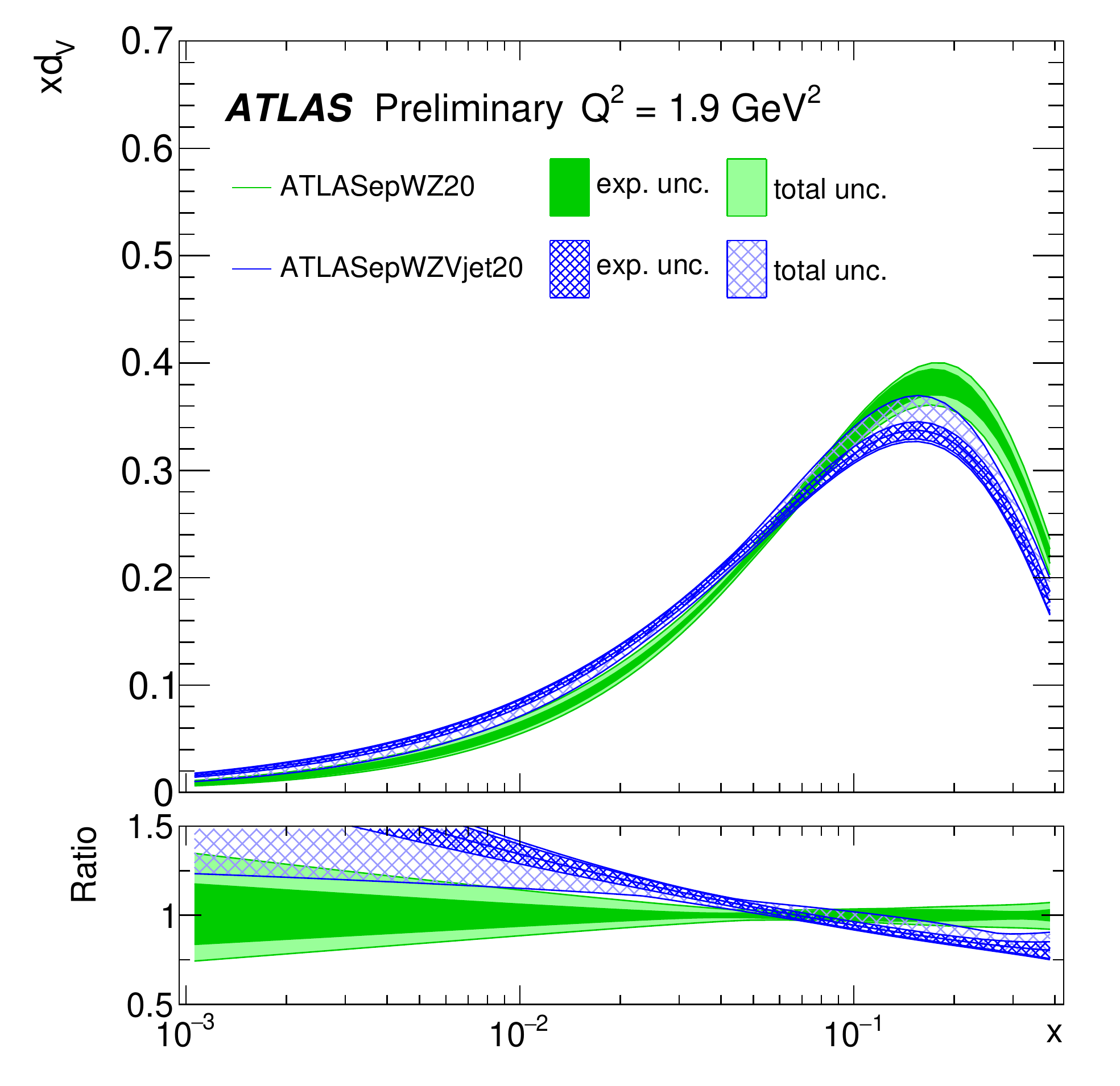}
\includegraphics[width=.9\linewidth]{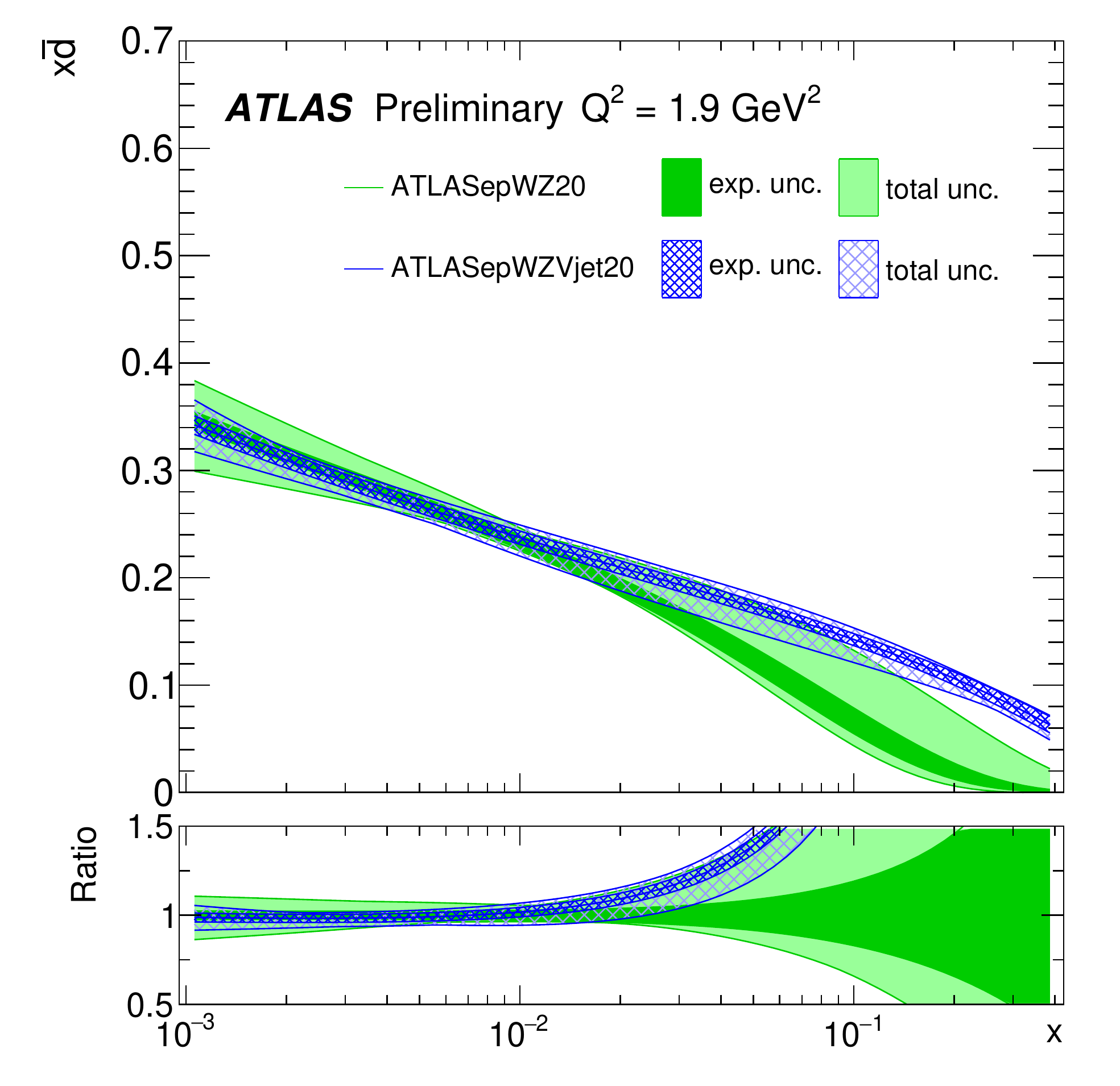}
\includegraphics[width=.9\linewidth]{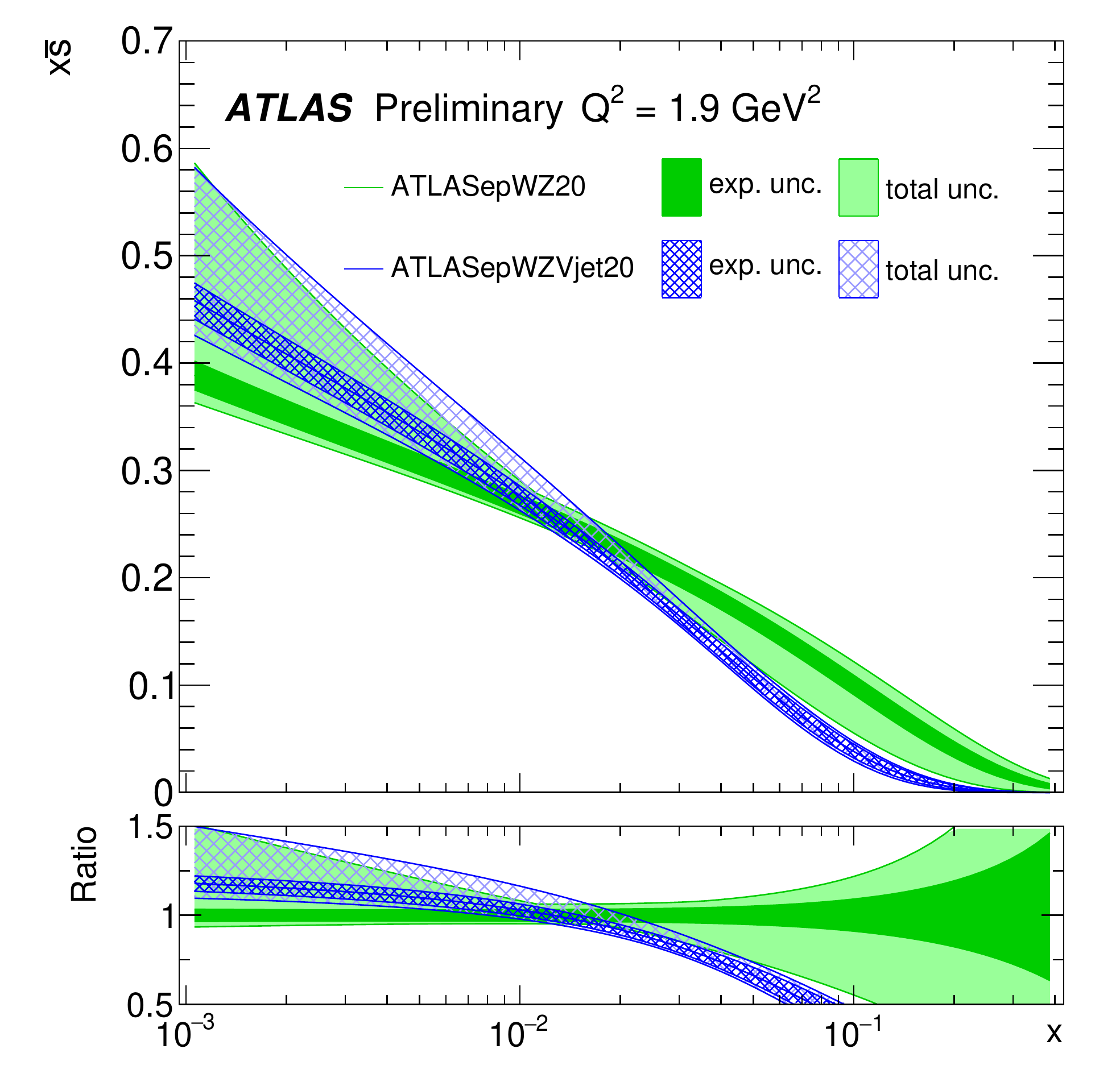}
	\caption{PDFs at the scale $Q^2$ = 1.9~GeV$^2$ as a function of Bjorken $x$ obtained for the $d_V$ (top), $\bar{d}$ (middle) and $\bar{s}$ (bottom) distributions when fitting $W$ + jets, $Z$ + jets, inclusive $W, Z$ and HERA data (ATLASepWZVjet20, blue bands), compared to a similar fit without $V$ + jets data (ATLASepWZ20, green bands). Inner error bands indicate the experimental uncertainty, while outer error bands indicate the total uncertainty, including parametrisation and model uncertainties. The ratio of each to the ATLASepWZ20 PDF set is displayed in the bottom panel in each case. These plots are taken from Ref.~\cite{CONF_2020}.}
\label{fig:VJetsPDFs}
\end{figure}
\begin{figure}[h!]
\centering
\includegraphics[width=.9\linewidth]{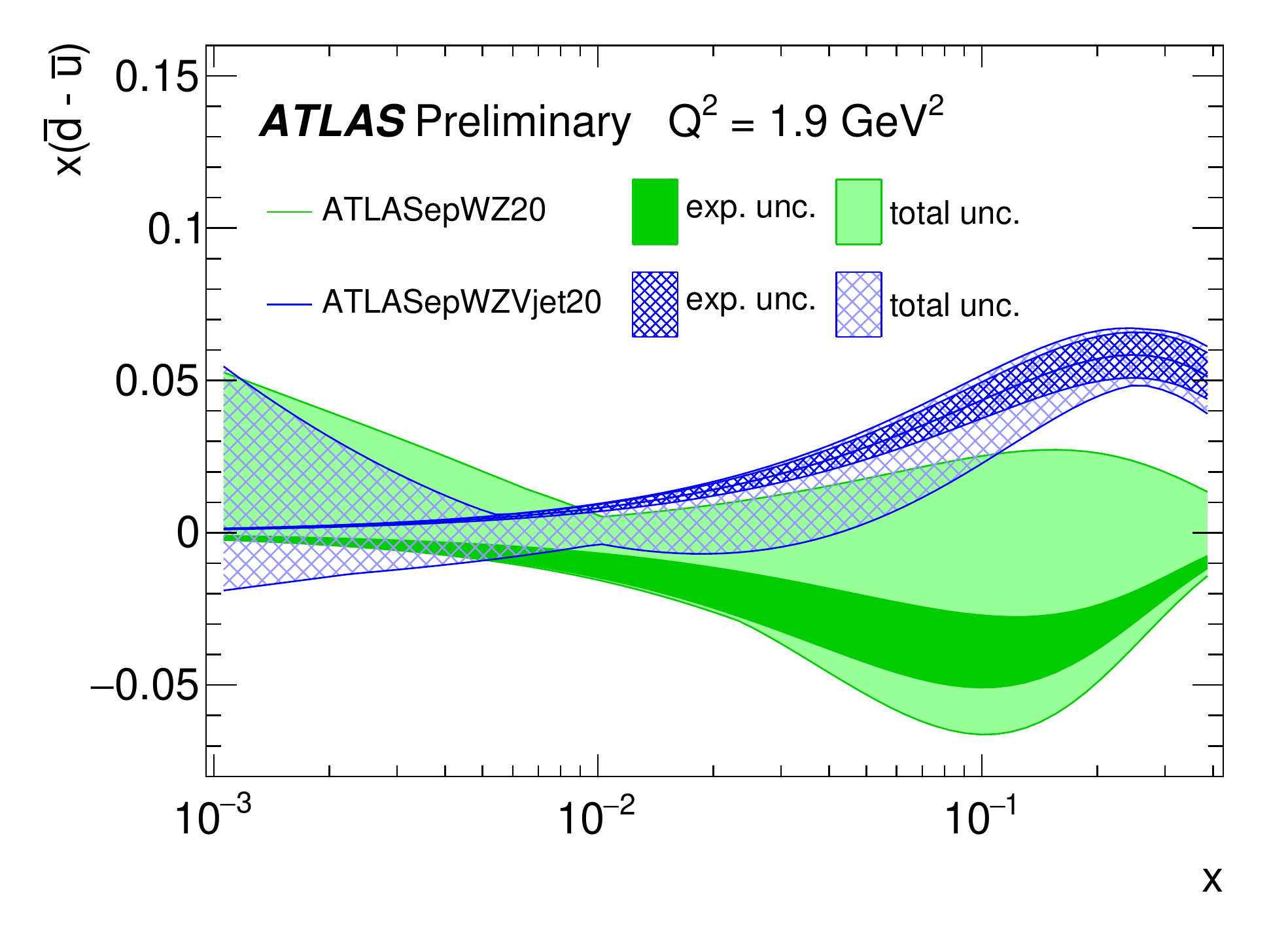}
\caption{The $x(\bar{d} - \bar{u})$ distribution evaluated at $Q^2$ = 1.9~GeV$^2$ as a function of Bjorken $x$ extracted from the ATLASepWZ20 (green) and ATLASepWZVjet20 (blue) fits with experimental and total uncertainties plotted separately. This plot is taken from Ref.~\cite{CONF_2020}.}
\label{fig:dbarubarWJets}
\end{figure}
\begin{figure}
\centering
\includegraphics[width=.9\linewidth]{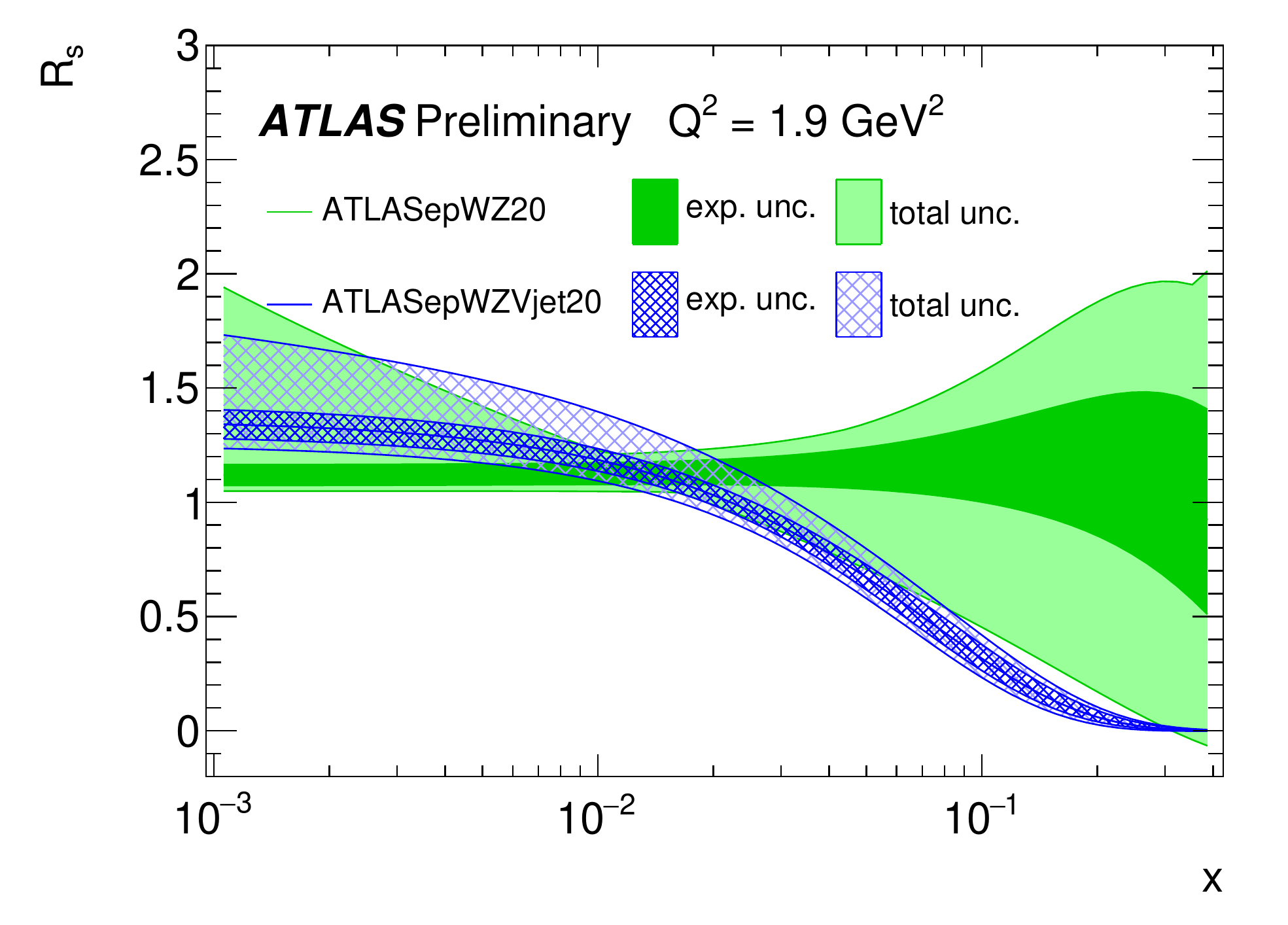}
\caption{The $R_{s}$ distribution, evaluated at $Q^2$ = 1.9~GeV$^2$, extracted from the ATLASepWZ20 (green) and ATLASepWZVjet20 (blue) fits with experimental and total uncertainties plotted separately. This plot is taken from Ref.~\cite{CONF_2020}.}
  \label{fig:RsWJets}
\end{figure}

Apart from experimental uncertainties evaluated at 68\% confidence level (CL), additional uncertainties in the PDFs are estimated and classified as model and parametrisation uncertainties.\\
Model uncertainties comprise variations of $m_c$, $m_b$, variations of the minimum $Q^2$ cut on data for entering the fit, and the starting scale at which the PDFs are parametrised, $Q_0^2$. The variation in charm-quark mass and starting scale are performed simultaneously to fulfil the condition $Q_0^2<m_c^2$, such that the charm PDF is calculated perturbatively. The parametrisation uncertainties are estimated through variations which include a single further parameter in the polynomial $P_{i}(x)$ or relaxed constraints of the low-$x$ sea quarks.\\
 The total uncertainty is the calculated as the sum in quadrature of the experimental, model and parametrisation uncertainties. Figure~\ref{fig:VJetsPDFs} show the ATLASepWZVjet20 PDFs overlaid with the ATLASepWZ20 PDFs, each evaluated at the starting scale $Q_0^2$, for comparison. The experimental and total uncertainties are displayed separately in each case.\\
The ATLASepWZVjet20 $x\bar{d}$ distribution is notably higher in the range $x \gtrsim 0.02$ compared to the ATLASepWZ20 fit. Similarly, the $x\bar{s}$ distribution of the ATLASepWZVjet20 fit in the same region is lower. Together, these changes allow for an increase in the $W^{+}$ cross-section, as depicted in Figure~\ref{fig:DataPlots1}. Additionally, the $d_v$ distribution is reduced at high $x$ and increased at low $x$. Finally, All the other distributions are similar between the two fits, hence they are not reported here.\\
The $x(\bar{d} - \bar{u})$ distribution as function of $x$ at $Q^2$ = 1.9~GeV$^2$ is shown in Figure~\ref{fig:dbarubarWJets}, with a comparison between ATLASepWZVjet20 and ATLASepWZ20 displaying the direct effect of the $V$ + jets data. The effect of the $V$ + jets data is to provide significant constraints to the total uncertainty at high-$x$, with an overall positive distribution of central values driven by the increase in the high-$x$ $\bar{d}$ distribution.\\
The fraction of the strange-quark density in the proton can be characterised by the quantity $R_s$, defined as the ratio
\begin{equation}
R_s = \frac{s + \bar{s}}{\bar{u}+\bar{d}}
\end{equation} 
which uses the sum of $\bar{u}$ and $\bar{d}$ as reference for the strange-sea density. The $R_s$ distribution plotted as a function of $x$ evaluated at $Q^2$ = 1.9~GeV$^2$ is shown in Figure~\ref{fig:RsWJets}, with a comparison between ATLASepWZVjet20 and ATLASepWZ20 showing the direct effect of the $V$ + jets data.
The effect of the $V$ + jets data is most significant in the kinematic region $x > 0.02$, where the uncertainty is significantly reduced.
Whereas the $R_s$ distribution of the ATLASepWZ20 PDFs maintained an unsuppressed strange distribution over a wide range in $x$, the ATLASepWZVjet20 PDFs exhibit an $R_s$ distribution falling from near-unity at $x\sim 0.01$ to approximately $0.5$ at $x=0.1$, driven by the increase in the high-$x$ $\bar{d}$ PDF and the complementary decrease in the high-$x$ $\bar{s}$ PDF shown in Figure~\ref{fig:VJetsPDFs}. At low $x \lesssim 0.023$ and $Q^2$ = 1.9~GeV$^2$, the fit with the $V$ + jets data maintains an unsuppressed strange-quark density compatible with the ATLASepWZ16 fit.

\subsection{Comparison to global PDFs}
\begin{figure}[t!]
\centering
\includegraphics[width=.9\linewidth]{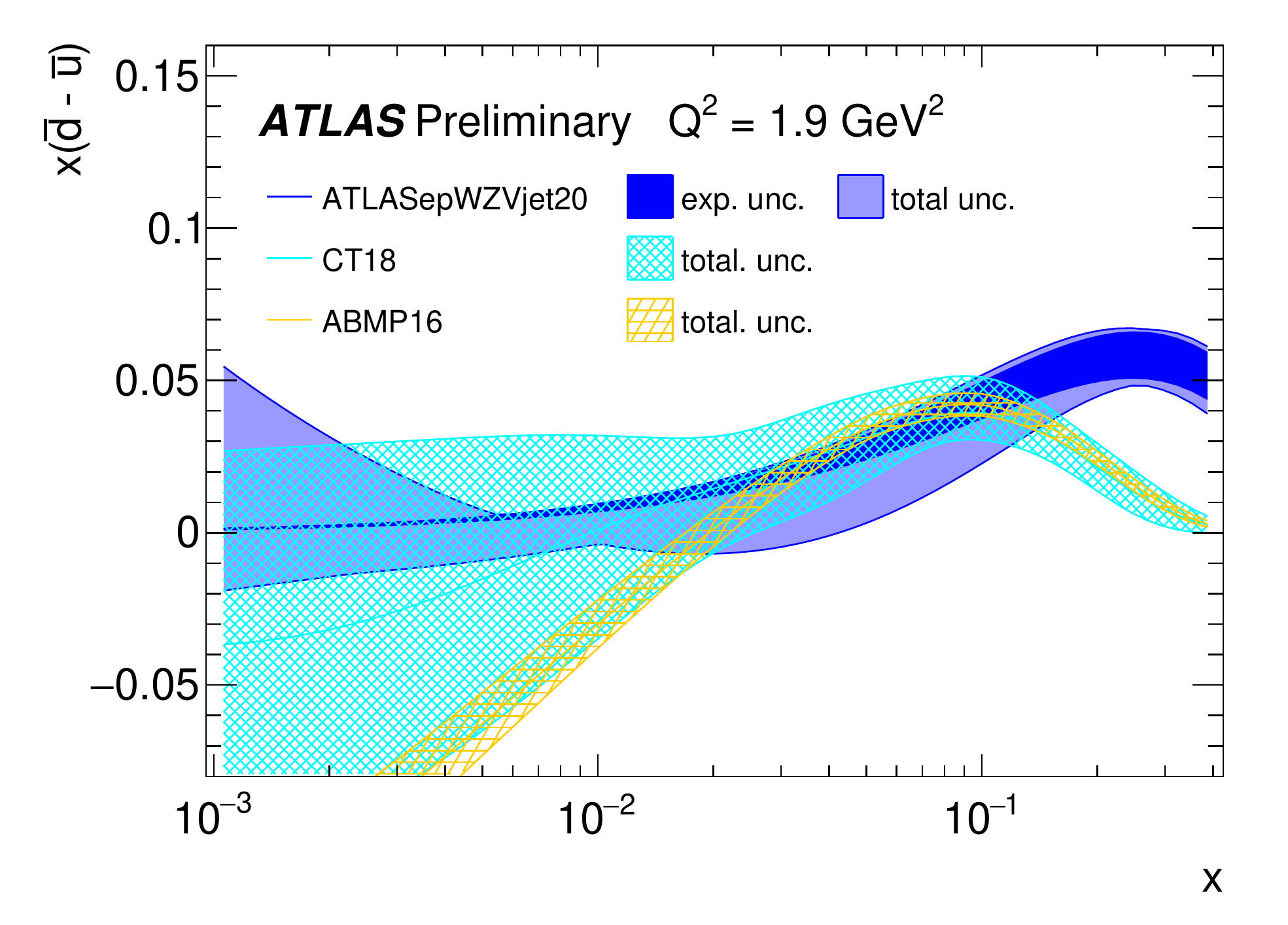}
\includegraphics[width=.9\linewidth]{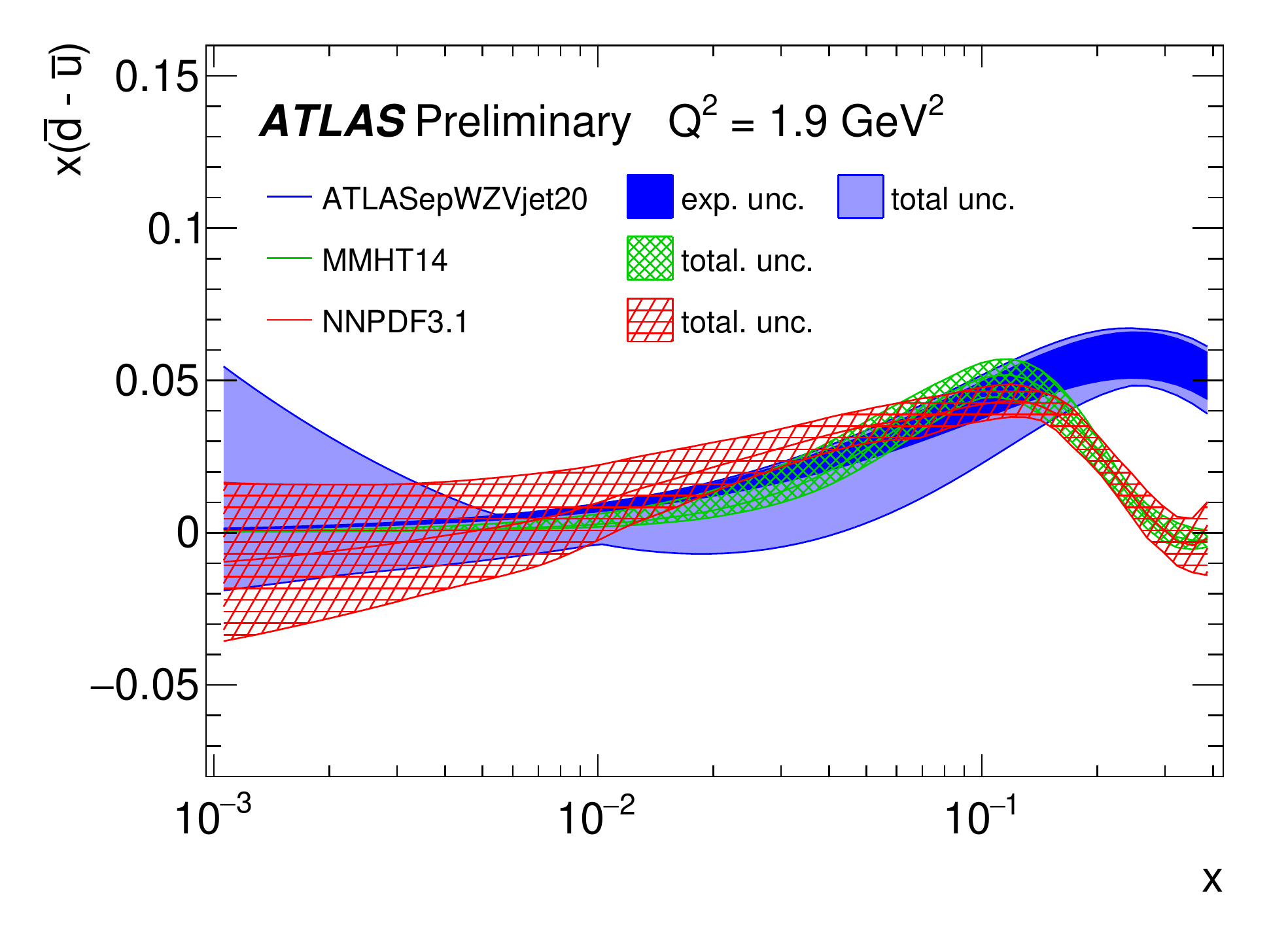}
\caption{The $x(\bar{d}-\bar{u})$ distribution evaluated at $Q^2$=1.9~GeV$^2$ as a function of Bjorken $x$, for the ATLASepWZVjet20 PDF set in comparison to global PDFs ABMP16 and CT18 (top), and MMHT14 and NNPDF3.1 (bottom). The experimental and total uncertainty bands are plotted separately for the ATLASepWZVjet20 result. Each global PDF set is taken at $\alpha_s(m_Z) = 0.1180$ except for ABMP16 which uses the fitted value $\alpha_s(m_Z) = 0.1147$. All global PDF uncertainty bands are at 68\% confidence level, evaluated for the CT18 PDFs through scaling by 1.645 as recommended by the PDF4LHC group \cite{Butterworth:2015oua}. These plots are taken from Ref.~\cite{CONF_2020}.}
\label{fig:dbarubarSummary}
\end{figure}
\begin{figure}
\centering
\includegraphics[width=.9\linewidth]{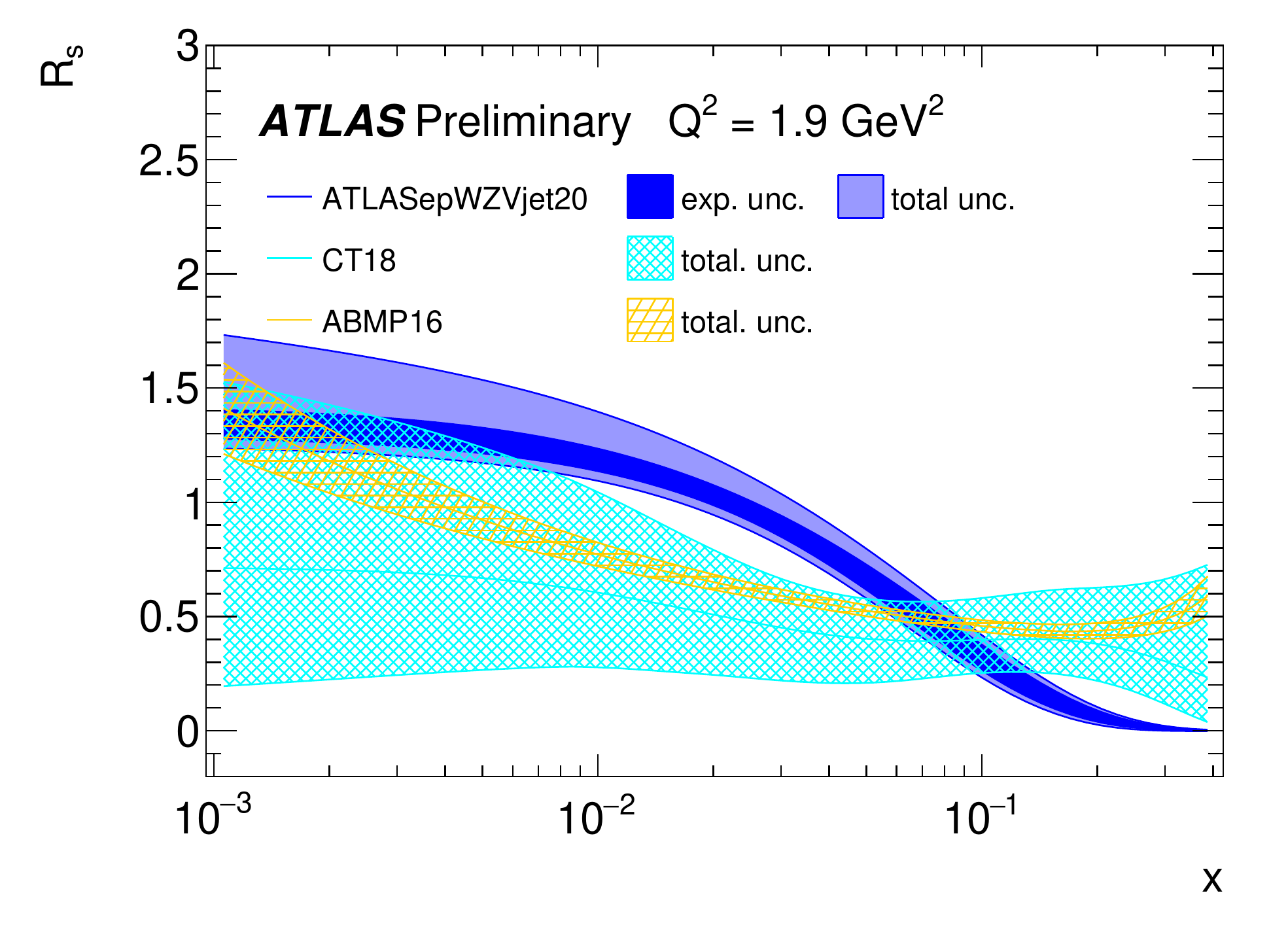}
\includegraphics[width=.9\linewidth]{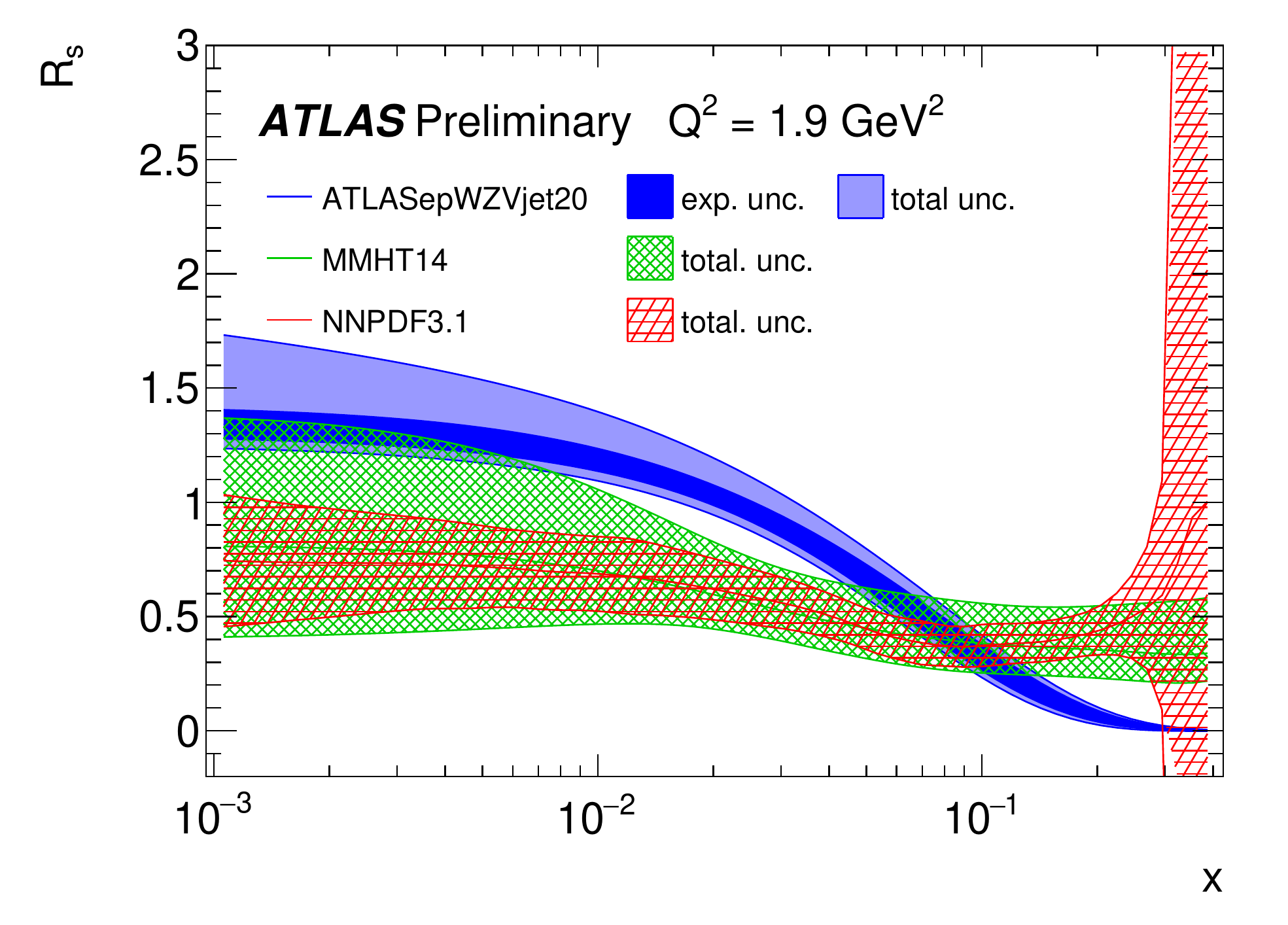}
\caption{The $R_s = (s + \bar{s})/(\bar{u}+\bar{d})$ distribution evaluated at $Q^2$ = 1.9~GeV$^2$ as a function of Bjorken $x$, for the ATLASepWZVjet20 PDF set in comparison to global PDFs ABMP16 and CT18 (top), and MMHT14 and NNPDF3.1 (bottom)~\cite{ABMP16,CT18,MMHT14,NNPDF31}. The experimental and total uncertainty bands are plotted separately for the ATLASepWZVjet20 results. Each global PDF set is taken at $\alpha_s(m_Z) = 0.1180$ except for ABMP16 which uses the fitted value $\alpha_s(m_Z) = 0.1147$. All global PDF uncertainty bands are at 68\% confidence level, evaluated for the CT18 PDFs through scaling by 1.645 as recommended by the PDF4LHC group \cite{Butterworth:2015oua}. These plots are taken from Ref.~\cite{CONF_2020}.}
\label{fig:RsSummaryFull}
\end{figure}
\begin{figure}
\centering
\includegraphics[width=.9\linewidth]{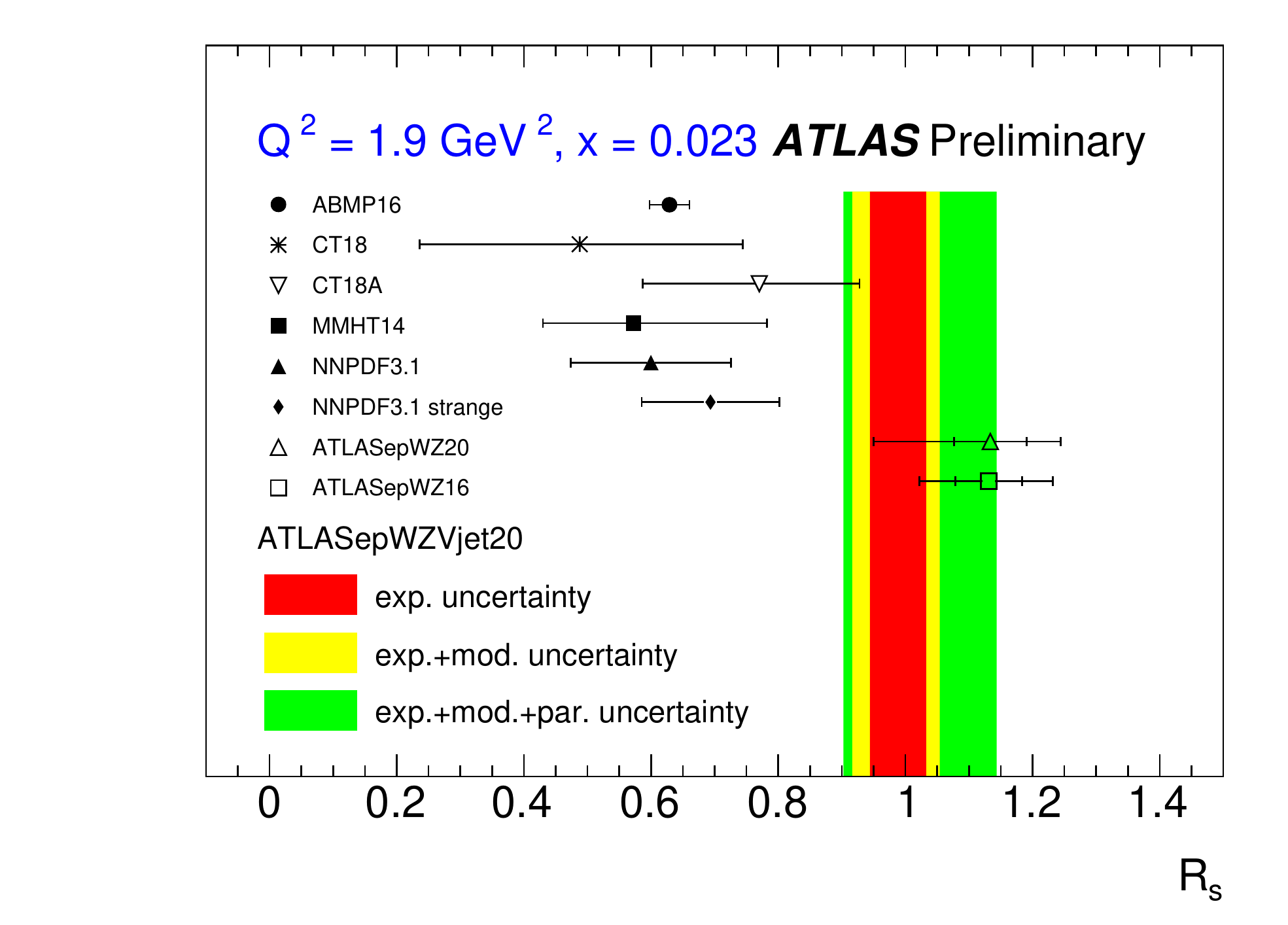}
\caption{Summary plots of $R_s$ evaluated at $x=0.023$ and $Q^2$ = 1.9~GeV$^2$, for the ATLASepWZVjet20 PDF set in comparison to global PDFs~\cite{CT18,MMHT14,NNPDF31,ABMP16,NNPDFStrange}, and the ATLASepWZ16 and ATLASepWZ20 sets. The experimental, model and parametrisation uncertainty bands are plotted separately for the ATLASepWZVjet20 results. Each global PDF set is taken at $\alpha_s(m_Z) = 0.1180$ except for ABMP16 which uses the fitted value $\alpha_s(m_Z) = 0.1147$. All uncertainty bands are at 68\% confidence level, evaluated for the CT18 PDFs through scaling by 1.645 as recommended by the PDF4LHC group \cite{Butterworth:2015oua}. This plot is taken from Ref.~\cite{CONF_2020}.}
  \label{fig:RsSummary}
\end{figure}

In Figure~\ref{fig:dbarubarSummary} the extracted $x(\bar{d}-\bar{s})$ distribution at $Q^2$ = 1.9~GeV$^2$ is shown in comparison to the results of the latest global PDF sets ABMP16~\cite{ABMP16}, CT18~\cite{CT18}, MMHT14~\cite{MMHT14} and NNPDF3.1~\cite{NNPDF31}.
The ATLASepWZVjet20 PDF set shows consistency with these global PDF sets up to $x \sim 0.1$, but deviates in the range $0.1 < x < 0.3$ where the $W$ + jets and $Z$ + jets data are most sensitive and demonstrate a preference for a higher $x\bar{d}$ distribution as discussed in Section~\ref{subsec:DataFits}.\\
The ATLASepWZVjet20 $R_s$ distribution over a range in $x$ is shown in Figure~\ref{fig:RsSummaryFull} in comparison to the above-mentioned global PDF sets. Tension between the ATLASepWZVjet20 fit and the global analyses is reduced compared to the ATLASepWZ16 and ATLASepWZ20 PDF sets, but persists to multiple standard deviations in the range $10^{-2} \lesssim x \lesssim 10^{-1}$. This is highlighted in summary plots of $R_s$ evaluated at $x=0.023$, $Q^2$ = 1.9~GeV$^2$ in Figure~\ref{fig:RsSummary}. However, better agreement is observed with the CT18A PDF set~\cite{CT18}, which includes the data used in the CT18 fit with the addition of ATLAS 7 TeV data. Furthermore, despite the inclusion of some additional data including the full ATLAS 7~TeV data set, tension remains with the recent update of the NNPDF3.1, labelled NNPDF3.1\_strange~\cite{NNPDFStrange}.
At high $x \gtrsim 0.02$, the $R_s$ distribution of the ATLASepWZVjet20 fit falls to zero, at a steeper rate than that of the global analyses.

\section{Conclusion}
This proceeding presents the impact of the ATLAS 8 TeV $V$ + jets data on PDFs, resulting in a new ATLASepWZVjet20 PDF set~\cite{CONF_2020}.
These data were fitted together with the data sets used for the previous ATLASepWZ16 fit~\cite{STDM-2012-20}, i.e. the full combined inclusive data set from HERA and the ATLAS inclusive $W$ and $Z$ production data recorded at $\sqrt{s}$ = 7 TeV at the LHC~\cite{LHC}. For the new ATLASepWZVjet20 PDF set, all significant systematic correlations between data sets were considered. \\
The resulting PDF set is similar to the ATLASepWZ16 set for the up-type quarks and gluon. The down and strange sea-quark distributions exhibit significantly smaller experimental and parametrisation uncertainties at high Bjorken $x$. As a result, the ratio of the strange to light quarks, $R_s$, is better constrained, falling more steeply at high $x$. The $x(\bar{d} - \bar{u})$ difference is positive, in better agreement with the global PDF analyses up to $x \sim 0.1$ but differing at higher values by up to two standard deviations.  At low $x \lesssim 0.023$, the fit shows consistency with an unsuppressed strange PDF as observed in the ATLASepWZ16 PDF set, while maintaining a positive $x(\bar{d}-\bar{u})$ distribution at high $x$.

\end{document}